\input harvmac.tex
\vskip 1.5in
\Title{\vbox{\baselineskip12pt 
\hbox to \hsize{\hfill}
\hbox to \hsize{\hfill  CTP-SCU/2014003 }}}
{\vbox{
	\centerline{\hbox{New Analytic Solutions in String Field Theory:
		}}\vskip 5pt
        \centerline{\hbox{Towards Collective Higher Spin Vacuum 
		}} } }
\centerline{Dimitri Polyakov$^{}$\footnote{$^{\dagger}$}
{polyakov@sogang.ac.kr ;
twistorstring@gmail.com
}}
\medskip
\centerline{\it Center for Theoretical Physics $^{(1)}$}
\centerline{\it College of Physical Science and Technology}
\centerline{\it Sichuan University, Chengdu 610064, China}
\centerline{\it }
\centerline{\it Center for Quantum Space-Time (CQUeST)$^{(2)}$}
\centerline{\it Sogang University}
\centerline{\it Seoul 121-742, Korea}
\centerline{\it }
\centerline{\it Institute for Information Transmission Problems (IITP)$^{(3)}$}
\centerline{\it Bolshoi Karetny per. 19/1}
\centerline{\it 127994 Moscow, Russia}
\vskip .3in

\centerline {\bf Abstract}

 We construct analytic solutions in cubic open superstring field theory at higher superconformal
ghost numbers.The solutions are the pure ghost ones,
given by combinations of Bell
polynomials of bosonized superconformal ghost fields 
multiplied by exponents of the bosonized ghosts.
Based on the structure of the solutions, we conjecture them 
to describe the ghost part of collective vacuum
for higher spin modes in open string theory.

\Date{August 2014}

\vfill\eject
\lref\fronsdalsit{C. Fronsdal, Phys.Rev. D20 (1979) 848-856}
\lref\fvf{E.S. Fradkin, M.A. Vasiliev, Nucl. Phys. B 291, 141 (1987)}
\lref\vcubic{M. A. Vasiliev, Nucl. Phys. B862 (2012) 341-408}
\lref\vmaf{M. A. Vasiliev, Sov. J. Nucl. Phys. 32 (1980) 439,
Yad. Fiz. 32 (1980) 855}
\lref\vmas{V. E. Lopatin and M. A. Vasiliev, Mod. Phys. Lett. A 3 (1988) 257}
\lref\vmulti{M. Vasiliev, Class.Quant.Grav. 30 (2013) 104006}
\lref\sagnottinew{A. Sagnotti, arXiv:1112.4285, J.Phys A46 (2013) 214006}
\lref\taronnao{M. Taronna, arXiv:1005.3061}
\lref\taronnas{A. Sagnotti, M. Taronna, arXiv:1006.5242 ,
Nucl.Phys.B842:299-361,2011}
\lref\hsatwo{E.S. Fradkin and M.A. Vasiliev, Mod. Phys. Lett. A 3 (1988) 2983}
\lref\fronsdal{C. Fronsdal, Phys. Rev. D18 (1978) 3624}
\lref\fradkin{E. Fradkin, M. Vasiliev, Phys. Lett. B189 (1987) 89}
\lref\klebanov{ I. Klebanov, A. M. Polyakov,
xPhys.Lett.B550 (2002) 213-219}
\lref\spinself{D. Polyakov, Phys.Rev.D82:066005,2010}
\lref\svv{E.D. Skvortsov, M.A. Vasiliev,
Nucl. Phys.B 756 (2006)117}
\lref\selfsw{D. Polyakov, J. Phys. A46 (2013) 214012}
\lref\szw{A. Sen, B. Zwiebach, JHEP 0003 (2000) 002}
\lref\selfff{D. Polyakov,  Phys.Rev. D89 (2014) 026010}
\lref\klebf{S. Giombi, I. Klebanov, arXiv:1308.2337}
\lref\schnablf{M. Schnabl, Adv.Theor.Math.Phys. 10 (2006) 433-501}
\lref\schnabls{T. Erler, M.  Schnabl,JHEP 0910 (2009) 066}
\lref\schnablt{M. Kroyter, Y. Okawa, M. Schnabl, S. Torii, B. Zwiebach, JHEP 1203 (2012) 030}
\lref\kroyter{M. Kroyter, JHEP 1103 (2011) 081}
\lref\berkf{N. Berkovits, A. Sen, B. Zwiebach,  Nucl.Phys. B587 (2000) 147-178}
\lref\berks{N. Berkovits, Nucl. Phys. B450 (1995) 90}
\lref\berkt{N. Berkovits,  JHEP 0004 (2000) 022}
\lref\iaf{I. Arefeva, P. Medvedev, A. Zubarev, Mod. Phys. Lett. A6, 949
(1991)}
\lref\ias{I. Arefeva, A. Zubarev,  Mod.Phys.Lett. A8 (1993) 1469-1476}
\lref\iat{I. Arefeva, P. Medvedev, A. Zubarev, Nucl.Phys. B341 (1990) 464-498}
\lref\discf{I. Klebanov, A. Polyakov,  Mod.Phys.Lett. A6 (1991) 3273-3281}
\lref\discs{E. Witten, Nucl.Phys. B373 (1992) 187-213}
\lref\disct{E. Fradkin, V. Linetsky, Mod.Phys.Lett. A4 (1989) 2635-2647}
\lref\discft{E. Fradkin, V. Linetsky, Mod.Phys.Lett. A4 (1989) 2649-2665}
\lref\discself{D. Polyakov, Int.J.Mod.Phys. A22 (2007) 1375-1394}
\lref\rastelli{L. Rastelli, B. Zwiebach, JHEP 0109 (2001) 038}
\lref\witsft{E. Witten, Nucl.Phys. B268 (1986) 253}
\lref\witsfts{E. Witten, Phys.Rev. D46 (1992) 5467-5473}
\lref\yost{C. Preitschopf, C. Thorn and S. Yost, Nucl. Phys. B337 (1990) 363}
\lref\erler{T. Erler, JHEP 0801:013, (2008)}
\lref\lbf{L. Bonora, C. Maccaferri, P. Prester, JHEP 0401 (2004) 038 }
\lref\lbs{L.  Bonora, C. Maccaferri, R. Scherer  Santos, D. Tolla,
Nucl.Phys. B715 (2005) 413-439}
\lref\lbt{L. Bonora, C. Maccaferri, D. Tolla, JHEP 1111 (2011) 107}
\lref\barsf{I. Bars, Phys.Rev. D66 (2002) 066003}
\lref\barss{I. Bars, I. Kishimoto, Y. Matsuo, Phys.Rev. D67 (2003) 126007}
\lref\bellf{E. T. Bell,  Annals of Mathematics 29 (1/4): 38–46}
\lref\bells{G. E. Andrews, ``The Theory of Partitions'' Cambridge University Press ISBN 0-521-63766-X.
(1998)}
\lref\bellt{K. Boyadzhiev, Abstract and Applied Analysis 2009: Article ID 168672}

\centerline{\bf  1. Introduction}
 String theory is known to be a powerful tool to approach problems
of describing consistently interacting higher spin (HS) field theories,
as well as  higher spin holography.
In string theory, the higher spin modes appear naturally
as vertex operators and the symmetries of the higher spin algebra
are realized in terms of the operator algebras of these vertices 
~{\fradkin, \hsatwo, \vmas, \fradkin, \taronnas, \sagnottinew, \spinself}.
As the on-shell constraints and the symmetry transformations 
on higher spin fields in space-time follow from the BRST conditions 
on the corresponding vertex operators in open or closed string theory,
the $N$-point correlation functions of higher spin vertex 
operators also define the gauge-invariant HS interactions  and, in the AdS
case, the  holographic couplings in dual CFT ~{\klebanov}
Unfortunately, however, the on-shell  string theory is background-dependent and
it is generally hard to approach string theory in AdS space beyond
semiclassical limit.
On the other hand, open string field theory (OSFT) is currently our best hope
to advance towards background independent formulation of strings, with
the OSFT equations of motion in the cubic-like
theory formally reminiscent of the relations
for the master fields in the Vasiliev's equations in the unfolding 
formalism for higher spins -  with the star products naturally
appearing in the both theories
 ~{\witsft, \witsfts, \yost, \iat, \ias, \hsatwo, \berks,
\berkf, \berkt, \rastelli, \lbf, \lbs, \lbt, \barsf, \barss }
At the same time, the form of the vertex operators in RNS string theory,
describing the higher spin gauge fields in the frame-like formalism ~{\selfff}
already carries a strong hint on their relevance to 
 background independence and emergent $AdS$ geometry.
Namely, consider open string vertex operators for Vasiliev type
two-row higher  spin gauge fields
$\Omega_m^{a_1...a_{s-1}|b_1...b_t}(x)\equiv\Omega^{s-1|t}(x)(0\leq{t}\leq{s-1})$
~{\vmas, \vmaf, \hsatwo, \svv}
where $m$ is the curved $d$-dimensional space index and
 $a,b$ indices  (corresponding to rows of lengths $s-1$ and $t$)
 label $d$-dimensional tangent space.
In case of $t=s-3$ the expression for the spin $s$ operator
particularly simplifies and is given by:
\eqn\grav{\eqalign{V_{s-1|s-3}(p)
\equiv
\Omega_m^{a_1...a_{s-1}|b_1...b_{s-3}}(p)V^m_{a_1...a_{s-1}|b_1...b_{s-3}}(p)
\cr
=
\Omega_m^{a_1...a_{s-1}|b_1...b_{s-3}}(p)\oint{dz}e^{-s\phi}\psi^m\partial
\psi_{b_1}\partial^2\psi_{b_2}...\partial^{s-3}\psi_{b_{s-3}}
\partial{X_{a_1}}...\partial{X_{a_{s-2}}}e^{ipX}}}
at minimal negative picture $-s$.
The manifest expressions for the spin $s$
operators with $0\leq{t}<{s-3}$ are generally more complicated,
however, at their  canonical pictures
equal to $-2s+t+3$,
 they can be related to the operator $V^m_{a_1...a_{s-1}|b_1...b_{s-3}}$
(1) by
\eqn\grav{\eqalign{
:\Gamma^{s-t-3}\Omega_m^{a_1...a_{s-1}|b_1...b_t}(p)V^m_{a_1...a_{s-1}|b_1...b_t}:(p)
=\Omega_m^{a_1...a_{s-1}|b_1...b_{s-3}}(p)V^m_{a_1...a_{s-1}|b_1...b_{s-3}}(p)}}
where $\Gamma=:e^\phi{G}:$ is the picture-changing operator
satisfying $:\Gamma^m\Gamma^n:=:\Gamma^{m+n}:+\lbrace{Q_{brst}},...\rbrace$,
$G$ is the full matter$+$ ghost worlsheet supercurrent
and

$:\Gamma^n:\sim:{e^{n\phi}}G\partial{G}...\partial^{n-1}G:$
The operator identity (2)  particularly entails a set of generalized
torsion zero constraints relating the space-time extra fields in the frame-like
formalism for the higher spins ~{\selfff}:
\eqn\grav{\eqalign{\Omega{s-1|s-3}(x)\sim\partial^{s-3-t}\Omega^{s-1|s-3}(x)}}
Although  the canonical pictures for the $V^{s-1|t}$
(defined by the singularity order in the asymtotic behavior of the
supermoduli approaching the insertion  point of a vertex operator)
are different for various $t$ values, this shouldn't be confused
with ghost cohomology ranks  which are the same for 
all the vertex operators of the extra fields with given $s$ and only depend
on the spin value $s$; that is, all the operators
for the frame-like fields of spin $s\geq{3}$ are the elements
of $H_{-s}\sim{H_{s-2}}$ ~{\spinself, \selfff}.
In the leading order, the low-energy equations of motion
for $\Omega^{s-1|t}$ extra fields are defined 
by the Weyl  invariance constraints on their vertex operators.
Naively, since the operators (1) are massless and originally are defined in 
flat space-time, one would expect the low-energy equations
of motions to be given simply by $\beta_\Omega{\sim}p^2\Omega^{s-1|t}(p)=0$ in
the leading order.
However, as was  explained in ~{\selfff}, due to nontrivial ghost
dependence of the operators in  $H_{-s}\sim{H_{s-2}}$ there are
anomalous ``mass-like'' terms appearing in the low-energy
equations of motion which, in the leading order, are actually given by
\eqn\grav{\eqalign{\beta_m^{a_1...a_{s-1}}=
-p^2\Omega_m^{a_1...a_{s-1}|b_1...b_t}(p)
+
{\Sigma_1}(a_1|a_2,...a_{s-1})p_dp^{a_1}\Omega_m^{a_2...a_{s-1}d|b_1...b_t}
\cr
-{1\over2}\Sigma_2(a_{s-2},a_{s-1}|a_1,...,a_{s-3})p^{a_{s-1}}p^{a_{s-2}}
(\Omega_m^\prime)^{a_1...a_{s-3}|b_1...b_t}
-4(s-1)\Omega_m^{a_1...a_{s-1}|b_1...b_t}}}
where $\Sigma_{1,2}$ are  the Fronsdal's symmetrization
operators ~{\fronsdalsit} and the prime stands for taking a trace over a couple
of indices.  For $t=0$,
these leading order equations of motion coincide with those
for the Fronsdal's massless higher spin fields in $AdS$ background,
polarized along the boundary  of $AdS$,
with the ``mass-like'' term related to the covariant derivative
of the Laplacian in $AdS$.
This leading order calculation already provides an example how curved
geometrical backgrounds (in this case, $AdS$) typically 
appear in calculations involving
vertex operators with nontrivial ghost dependence.
Unfortunately, the calculation performed in ~{\selfff} is hard  to 
generalize beyond the leading order, particularly
because the operators with different $t$ values mix with each other
at the level of $3$-point functions and beyond, so the 
straightforward $\beta$-function calculations become cumbersome
and practically insurmountable.
Given the fact that the conformal $\beta$-function is essentially
an off-shell object and , in case of 
 the higher spin operators, demonstrates the background change,
this clearly makes a string field theory, extended
to higher superconformal ghost numbers, a natural framework
to approach the problem of higher spin interactions from string-theoretic
point of view.
First of all, recall that the background independence in cubic string field
theory implies that
the equations of motion:
\eqn\lowen{Q\Psi+\Psi\star\Psi=0}
are invariant under the shift $\Psi\rightarrow\Psi+\Psi_0$
where $\Psi_0$ is a solution of (5), provided that the BRST charge is
shifted according to:
$Q\rightarrow{Q}+\Psi_0$, implying
$Q\Psi\rightarrow{\tilde{Q}}\Psi={Q\Psi}+\Psi_0\star\Psi+\Psi\star\Psi_0$
for any string field $\Psi$. It is important that, in this approach,
the new BRST charge, ${\tilde{Q}}$ defines new BRST cohomology and string field theory around
certain new geometrical background, even though the original theory is 
often defined  around the flat vacuum (making it possible to compute
the off-shell correlators). For this reason, exact analytic solutions
in OSFT are crucial ingredients to  analyze the off-shell interactions
in various space-time geometries (such as AdS), holding a key to background independence.
Unfortunately, because of the complexity of the equations (5) very few concrete examples of the 
analytic OSFT solutions are known so far.
One remarkable example is of course the nonperturbative tachyon vacuum solution proposed by Schnabl 
~{\schnablf},
generalized in many subsequent importants papers ~{\schnabls, \schnablt, \erler} 
 and  particularly used to
prove the Sen's conjectures on tachyon condensation ~{\schnablf, \schnabls}
The solutions found by Schnabl ~{\schnablf} particularly used the basis of the wedge states
where the star product simplifies significantly ~{\rastelli, \schnablf}
Nevertheless, because of the complexity 
of the star product (7)  calculating the star products of string fields is generally an extremely
cumbersome and tedious problem. The reason for this complexity is that,
in terms of correlation functions in OSFT,  the star product
involves conformal transformations mapping the worldsheets of interacting strings
to wedges of a single disc. For example,
in case of $ N$ interacting strings  the transformation mapping
the worldsheet of the n-th string to the n-th wedge ($1\leq{n\leq{N}}$)
is given by
\eqn\grav{\eqalign{f_n^N(z)=e^{{{i\pi(n-1)}\over{N}}}({{1-iz}\over{1+iz}})^{{2\over{N}}}}},
and the star product of two string fields
is defined according to
\eqn\grav{\eqalign{<<\Phi;\Psi_1\star\Psi_2>>=
<f_1^3{\circ}\Phi(0)f_2^3\circ\Psi_1(0)f_3^3\circ\Psi_2(0)>}}
for an arbitrary string field $\Phi$
Here  $f\circ\Psi$ is conformal transformation acting on all the operators
(generically, off-shell) entering $\Psi$.
For primary fields $V_h$ of conformal dimension $h$, the transformation is simply given
by
\eqn\grav{\eqalign{f\circ{V_h}(0)=
({{df}\over{dz}})^hV_h(f(0))}}
and in case if descendents are given  by
 the derivatives of the primaries the transformation is given by differentiating
(8).
However, for generic non-primaries the global transformation laws are far more complicated and
cannot  be reduced to differentiating the combinations of (8)
(e.g. recall the simplest example - the global transformation law
of a stress tensor involving Schwarzian derivatives)
Therefore a generic string field transforms under (6) 
in a complicated manner that is hard to control, so 
straightforward  calculation of the
star product through  correlators is not accessible.
However, as we  point out in this work, there exists  a  substantially large class of operators
for which the global conformal transformations simplify significantly, having an elegant 
and compact form. This class includes Bell polynomials  of bosonized superconformal ghost fields
multiplied by exponents of the bosonized ghosts.These operators altogether form an invariant subspace under global
conformal transformations, making it possible to directly deduce the star product from
the correlation functions for the string fields restricted to this subspace.
In this paper we consider the cubic superstring field theory ~{\yost, \iaf, \ias}
We impose no restrictions  on superconformal ghost numbers of string field components,
allowing them to have any positive or negative pictures. We furthermore impose
weak cohomology condition (44) on the string field replacing the standard gauge fixing
(see below); this condition ensures that string field components at different
superconformal ghost numbers are not mixed by picture changing 
For this reason, there is no need for the midpoint insertions of inverse picture-changing
operators (unlike the cases when the ghost number and picture of string fields are fixed
or the model of democratic superstring field theory considered in ~{\kroyter})
In this paper we particularly propose an ansatz for an analytic solution in 
string field theory with the structure:
\eqn\grav{\eqalign{
\Psi=\Psi^{(+)}+\Psi^{(-)}
\cr
\Psi^{(+)}=\sum_{N=1}^\infty\sum_{n=0}^{N-2}\lambda_N^n{c}e^{\chi+N\phi}
B_n^{\lbrack\alpha_n,\beta_n,\gamma_n\rbrack}(\phi,\chi,\sigma)
\cr
\Psi^{(-)}=\sum_{N=1}^\infty\sum_{n=0}^{N-2}\lambda_N^n{c}e^{-(N+2)\phi}
B_n^{\lbrack\alpha_n,\beta_n,\gamma_n\rbrack}(\phi,\chi,\sigma)}}

The limits of summations over $n$ are related to the ghost cohomology 
constraints on $\Psi^{(+)}$ and $\Psi^{(-)}$ (see below).
The $\lambda_N^n$-coefficients in front of  $ce^{\chi+N\phi}$
and $ce^{-(N+2)\phi}$ are chosen  equal in order 
to preserve the isomorphism between negative and positive ghost cohomologies
$H_{n}\sim{H_{-(n+2)}}$ ~{\spinself}
and $\alpha_n,\beta
_n,\gamma_n$  are certain numbers
chosen to satisfy ghost cohomology constraints on $\Psi$ (see below).  The Bell polynomials
in bosonized ghost fields can be computed according to
\eqn\grav{\eqalign{B_n^{\lbrack\alpha_n,\beta_n,\gamma_n\rbrack}(\phi,\chi,\sigma)=
e^{-\alpha_n\phi-\beta_n\chi-\gamma_n\sigma}(z){{d^n}\over{dz^n}}e^{\alpha_n\phi+\beta_n\chi+\gamma_n\sigma}(z)
}}
 implying $x_k=\alpha_k\partial\phi+\beta_k\partial\chi+\gamma_k\partial\sigma$
in the formal definition (14), (15).
The analytic solution with the structure
(9) is then calculated in our work in terms of recurrence relations
satisfied by $\lambda_N^n$ coefficients:
\eqn\grav{\eqalign{
\lambda^n_{N}=\sum_{N_1,N_2=1}^{N_1+N_2=N-2}\sum_{n_1=0}^{N_1-1}\sum_{n_2=0}^{N_2-1}\rho_{N|N_1;N_2}^{n|n_1;n_2}
\lambda_{N_1}^{n_1}\lambda_{N_2}^{n_2}}}
Our purpose is to determine the coefficients:
{$\rho_{N|N_1,N_2}^{n|n_1;n_2}$} 
by  directly computing the star product, i.e. the relevant correlators
{
$$<<\Psi,Q\Psi>>=<\Psi(0)I\circ{Q\Psi}(0)>$$}
and 
{$$<<\Psi,\Psi\star\Psi>>
=<h{\circ}f_1^3\circ\Psi(0)h{\circ}f_2^3\circ\Psi(0)h{\circ}f_3^3\circ\Psi(0)>$$}
where
{$I(z)=-{1\over{z}}$}
and
\eqn\grav{\eqalign{
f_k^n(z)=e^{{i\pi(k-1)}\over{n}}({{1-iz}\over{1+iz}})^{2\over{n}}}}

maps the worldsheets of $n$ interacting strings  
 putting them together on a single disc and
\eqn\lowen{h(z)=-i{{z-1}\over{z+1}}}
maps this disc back to the halfplane.

 The rest of this paper is  organized as follows.
In section 2 we review basic facts about Bell polynomials and derive the global conformal
transformation rules for operators entering the proposed ansatz for the exact analytic solution.
In the the Section 3 we compute the relevant correlators in OSFT and derive 
the recursion relations for the coefficients defining the solution.
In the concluding section we discuss the relevance of the solution to higher spin algebras in $AdS$
and possible generalizations.

\centerline{\bf 2. Bell Polynomials and Global Conformal Transformations}

The standard definition of the complete Bell  polynomials ~{\bellf, \bells, \bellt}
$B_n(x_1,...x_n)$ is given by
\eqn\lowen{B_n(x_1,...x_n)=\sum_{k=1}^n{B_{n|k}}(x_1,...x_{n-k+1})}
where $B_{n|k}(x_1,...x_{n-k+1})$ are the partial Bell polynomials
defined according to
\eqn\grav{\eqalign{B_{n|k}(x_1,...x_{n-k+1})=
\sum_{p_1,...p_{n-k+1}}{{n!}\over{p_1!...p_{n-k+1}!}}x_1^{p_1}({{x_2}\over{2!}})^{p_2}
...({{x_{n-k+1}}\over{(n-k+1)!}})^{p_{n-k+1}}}}
with the sum taking over all the combinations of non-negative  $p_j$ satisfying
\eqn\grav{\eqalign{
\sum_{j=1}^{n-k+1}p_j=k\cr
\sum_{j=1}^{n-k+1}jp_j=n}}
In number theory,  Bell polynomials are known to satisfy a number of 
useful and beautiful
identities and properties, with some of them important for the  
calculations performed in this paper ~{\bellf, \bells}
Just to mention a couple of examples,
\eqn\grav{\eqalign{B_{n|k}(1,...1)=S(n,k)}}
is the second kind Stirling number 
and $B_{n|k}(0!,1!,...,(n-k)!)$ can be expressed in terms of
combinations of Bernoulli numbers
(note the appearance of Bernoulli numbers
in the analytic SFT solutions  describing the tachyonic vacuum ~{\schnablf}).
Also, given a Taylor's expansion of a  function 
$f(x)=\sum_n{{a_nx^n}\over{n!}}$
one has
$e^f(x)=\sum_n{B_n(a_1,..a_n)}
{{x^n}\over{n!}}$ in terms of formal series,
so e.g. vertex operators in string theory are typically
given by combinations of Bell polynomials in the expansion modes.
Note that the SFT ansatz (9) is actually  bilinear in Bell polynomials of
the ghost expansion modes.
If one identifies $x_n=\partial^n{\phi}(z)$, where $\phi(z)$ is some scalar field,
one obtains Bell polynomials in derivatives of $\phi$;
note that in the particular case $\phi(z){\sim}z^2$ this would reduce to Hermite polynomials in $z$.
Other useful objects to define are the Bell generators
\eqn\grav{\eqalign{H_n(y|x_1,...,x_n)=\sum_{k=1}^nB_{n|k}(x_1,...,x_{n-k+1})y^n}}
and more generally
\eqn\grav{\eqalign{G_n(y_1...y_n|x_1,...,x_n)=\sum_{k=1}^nB_{n|k}(x_1,...,x_{n-k+1})y_1,...,y_k}}

In the context of two-dimensional CFT, one can think of Bell polynomials as higher derivative
generalizations of the Schwarzian derivative, appearing in the global conformal transformation law
for the stress tensor.
That is, under $z\rightarrow{f(z)}$
one has
\eqn\grav{\eqalign{T(z)\rightarrow({{df}\over{dz}})^2T(f(z))+{c\over{12}}S(f(z))}}
where the Schwarzian derivative:
\eqn\grav{\eqalign{
S(f(z))=({{f^{\prime\prime}(z)}\over{f^\prime(z)}})^{\prime}-{1\over2}
({{f^{\prime\prime}(z)}\over{f^\prime(z)}})^2}}
can be expressed in terms of the second order Bell polynomials in the
log of $f^\prime$, with $x_k\equiv{{d^{k-1}}\over{dz^{k-1}}}log(f^\prime)$:
\eqn\grav{\eqalign{
S(f(z))=B_{2|1}(log(f^\prime),{{dlog(f^\prime)}\over{dz}})-{1\over2}
B_{2|2}(log(f^\prime))
\cr
\equiv{-2H_2(-{1\over2}|log(f^\prime))}
\equiv{-2}B_2(-{1\over2}log(f^\prime))
}}
where, for the sake of brevity, we adopt the notation:
\eqn\grav{\eqalign{B_n(g(x))
\equiv{B_n(\partial{g},...\partial^n{g})}
=B_n(x_1,...x_n)|_{x_k=\partial_x^k{g}(x);k=1,...,n}
}}
for any function $g(x)$.
This point is  of importance as the  higher order Bell polynomials will
naturally enter the global transformation law for the string fields of the type (9)
(see below).
We now turn to the question of  finding global conformal transformation
law for the OSFT ansatz (9).  Our strategy will be to find the infinitezimal
form of the transformation first and then to deduce the global transformation
by requiring it to reproduce the infinitezimal one while preserving its form under
the composition of two global transformations.
We start with the infinitezimal transformation following from the OPE 
of the stress tensor $T(z)$ with
$B_n^{\lbrack\alpha_n,\beta_n,\gamma_n\rbrack}e^{\lbrack\alpha.\beta\,\gamma\rbrack}$
where $\alpha,\alpha_n,....$ are  some numbers
and
$$e^{\lbrack\alpha.\beta\,\gamma\rbrack}\equiv{e^{\alpha\phi+\beta\chi+\gamma\sigma}}$$
in our notations.
 Consider the transformation of 
$B_n^{\lbrack\alpha_n,\beta_n,\gamma_n\rbrack}$ first.
It is calculated easily noting that
\eqn\grav{\eqalign{
B_n^{\lbrack\alpha_n,\beta_n,\gamma_n\rbrack}e^{\lbrack\alpha_n.\beta_n\,\gamma_n\rbrack}
=\partial^n{e^{\lbrack\alpha_n.\beta_n\,\gamma_n\rbrack}}}}
and
\eqn\lowen{T(z)e^{\lbrack\alpha_n.\beta_n\,\gamma_n\rbrack}(w)=
{{h^{\lbrack{\alpha_n,\beta_n,\gamma_n}\rbrack}}\over{(z-w)^2}}
+
{{{\partial}e^{\lbrack\alpha_n.\beta_n\,\gamma_n\rbrack}}\over{z-w}}+...}
where 
\eqn\lowen{
h^{\lbrack{\alpha_n,\beta_n,\gamma_n}\rbrack}={1\over2}(-\alpha^2+\beta^2+\gamma^2)
-\alpha-{\beta\over2}-{{3\gamma}\over2}}
stand for conformal dimensions of the exponents (with similar notations below).
One then easily computes:
\eqn\grav{\eqalign{
T(z)
B_n^{\lbrack\alpha_n,\beta_n,\gamma_n\rbrack}e^{\lbrack\alpha_n.\beta_n\,\gamma_n\rbrack}(w)
\cr
=\sum_{k=0}^{n+1}{{n!}\over{(n-k+1)!}}
{{kh^{\lbrack{\alpha_n,\beta_n,\gamma_n}\rbrack}+n-k+1}\over{(z-w)^{k+1}}}
B_{n-k+1}^{\lbrack\alpha_n,\beta_n,\gamma_n\rbrack}e^{\lbrack\alpha_n.\beta_n\,\gamma_n\rbrack}(w)
\cr
+...}}
To compute the variation of $B_n^{\lbrack\alpha_n,\beta_n,\gamma_n\rbrack}$ under infinitezimal
conformal transformation, we note that
\eqn\grav{\eqalign{
\delta_\epsilon(B_{n}^{\lbrack\alpha_n,\beta_n,\gamma_n\rbrack})e^{\lbrack\alpha_n.\beta_n\,\gamma_n\rbrack}
\cr
=
\delta_\epsilon(B_{n}^{\lbrack\alpha_n,\beta_n,\gamma_n\rbrack}e^{\lbrack\alpha_n.\beta_n\,\gamma_n\rbrack})
-
B_{n}^{\lbrack\alpha_n,\beta_n,\gamma_n\rbrack}\delta_\epsilon(e^{\lbrack\alpha_n.\beta_n\,\gamma_n\rbrack})
\cr
-overlap}}
with the overlap  contribution stemming from the leading order singularity
in the OPE of the quadratic part of the stress energy tensor with
one of $\partial\phi,\partial\chi,\partial\sigma$  coupling
with $B_n$ and another with the exponent.

To compute this contribution, one first has to calculate the OPE of $\partial\phi$ with
$B_{n}^{\lbrack\alpha_n,\beta_n,\gamma_n\rbrack}$  (analogously, for $\partial\chi$ and $\partial\sigma$)
This can be done directly, by using straightforward expression for $B_n$:

\eqn\grav{\eqalign{B_{n}^{\lbrack\alpha_n,\beta_n,\gamma_n\rbrack}=
\sum_{m=1}^n\sum_{n|p_1...p_m}{{{n!}\over{p_1!...p_m!q_{p_1}!...q_{p_m}!}}}
\prod_{j=1}^m(\alpha\partial^{p_j}\phi+\beta\partial^{p_j}\chi+\gamma\partial^{p_j}\sigma)
}}
where $n|p_1...p_m$ are the ordered partitions
of $n$ into m numbers $0<p_1\leq{p_2}...\leq{p_m}$
and $q_{p_j}$ are a multiplicities of $p_j$'s entering a partition.
Then the OPE is easily computed to give

\eqn\grav{\eqalign{\partial\phi(z)
B_{n}^{\lbrack\alpha_n,\beta_n,\gamma_n\rbrack}(w)=
-\alpha_n\sum_{k=1}^n{{n!}\over{(n-k)!}}{{B_{n-k}^{\lbrack\alpha_n,\beta_n,\gamma_n\rbrack}}\over{(z-w)^{k+1}}}
+O(z-w)^0}}

and similarly for $\partial\chi$ and $\partial\sigma$.
Using (30), the overlap contribution to (28) is  given by
\eqn\grav{\eqalign{
T(z)B_{n}^{\lbrack\alpha_n,\beta_n,\gamma_n\rbrack}(w)=(-\alpha^2+\beta^2+\gamma^2)\sum_{k=1}^n
{{B_{n-k}^{\lbrack\alpha_n,\beta_n,\gamma_n\rbrack}}\over{(z-w)^{k+2}}}}}
Combining (28) and (31),
the infinitezimal conformal transformation of 
$B_{n}^{\lbrack\alpha_n,\beta_n,\gamma_n\rbrack}$ is given by
\eqn\grav{\eqalign{\delta_\epsilon{B_{n}^{\lbrack\alpha_n,\beta_n,\gamma_n\rbrack}}
=\epsilon\partial{B_{n}^{\lbrack\alpha_n,\beta_n,\gamma_n\rbrack}}
+n\partial\epsilon{B_{n}^{\lbrack\alpha_n,\beta_n,\gamma_n\rbrack}}
\cr
+
\sum_{k=2}^{n+1}{{n!}\over{(n-k+1)!k!}}\partial^k\epsilon(z)
{{kh^{\lbrack{\alpha_n,\beta_n,\gamma_n}\rbrack}+n-k+1+(\alpha_n^2-\beta_n^2-\gamma_n^2)}}
B_{n-k+1}^{\lbrack\alpha_n,\beta_n,\gamma_n\rbrack}(z)}}
Finally, the infinitezimal conformal transformation of terms
entering the OSFT ansatz (9)
is  given by:
\eqn\grav{\eqalign{\delta_\epsilon
({B_{n}^{\lbrack\alpha_n,\beta_n,\gamma_n\rbrack}}
e^{\lbrack\alpha,\beta,\gamma\rbrack})
\delta_\epsilon({B_{n}^{\lbrack\alpha_n,\beta_n,\gamma_n\rbrack}})e^{\lbrack\alpha,\beta,\gamma\rbrack}
+{B_{n}^{\lbrack\alpha_n,\beta_n,\gamma_n\rbrack}}\delta_\epsilon(e^{\lbrack\alpha,\beta,\gamma\rbrack})
\cr
+overlap}}  with the overlap contribution in (33) stemming from the 
leading order  singularity in the OPE of the quadratic
part of $T(z)$ with 
${B_{n}^{\lbrack\alpha_n,\beta_n,\gamma_n\rbrack}}e^{\lbrack\alpha,\beta,\gamma\rbrack}$
with one of $\partial(\phi,\chi,\sigma)$ contracting with $B_n$  and another with
$e^{\lbrack\alpha,\beta,\gamma\rbrack}$ so the overall infinitezimal transformation 
of the string field components is

\eqn\grav{\eqalign{\delta_\epsilon({B_{n}^{\lbrack\alpha_n,\beta_n,\gamma_n\rbrack}}e^{\lbrack\alpha,\beta,\gamma\rbrack})
=\epsilon\partial({B_{n}^{\lbrack\alpha_n,\beta_n,\gamma_n\rbrack}}e^{\lbrack\alpha,\beta,\gamma\rbrack})
+\partial\epsilon(n+h^{\lbrack{\alpha,\beta,\gamma}\rbrack}){B_{n}^{\lbrack\alpha_n,\beta_n,\gamma_n\rbrack}}e^{\lbrack\alpha,\beta,\gamma\rbrack}
\cr
+
\sum_{k=2}^{n+1}{{n!}\over{(n-k+1)!k!}}\partial^k\epsilon(z)\lbrack
kh^{\lbrack{\alpha_n,\beta_n,\gamma_n}\rbrack}+n-k+1
\cr
+
(\alpha_n^2-\alpha_n\alpha-\beta_n^2+\beta_n\beta-\gamma_n^2+\gamma_n\gamma)
\rbrack
B_{n-k+1}^{\lbrack\alpha_n,\beta_n,\gamma_n\rbrack}e^{\lbrack\alpha,\beta,\gamma\rbrack}(z)}}

Given the infinitezimal transformation (34) with some effort one can deduce the correct form
of the global conformal transformation of the string field components (9)
under $z\rightarrow{f(z)}$ by requiring that:

1) it reproduces the transformation (34) for $f(z)=z+\epsilon(z)$

2) its form is preserved under the composition of two transformations

(as in the standard derivation of the transformation law for the stress tensor, leading
to the appearance of Schwarzian derivative, which is simply the second order
Bell polynomial in the $log$ of $f^\prime$ (22).)

Regarding the first condition, note that
\eqn\grav{\eqalign{B_n(log(f^\prime(z)))|_{f(z)=z+\epsilon(z)}=\partial^n\epsilon(z)+O(\epsilon^2)}}
Regarding the second,  note the binomial property of $B_n(f)$:
\eqn\grav{\eqalign{B_n(f(x)+g(x))\equiv
B_n(\partial(f+g),...,\partial^n(f+g))=
\sum_{k=0}^n{{n!}\over{k!(n-k)!}}B_k(f)B_{n-k}(g)}}
which obviously follows 
from the chain rule applied to the derivative $\partial_z^n{e^{f(z)+g(z)}}$
and , finally, ${{d}\over{dz}}log(g(f(z)))=
log{(g^\prime(f))}+log{(f^\prime(z))}$ for the composition of two conformal transformations
 $f$  and $g$.
This altogether fixes the form of the global conformal transformation according to

\eqn\grav{\eqalign{
B_{n}^{\lbrack\alpha_n,\beta_n,\gamma_n,\rbrack}e^{\lbrack\alpha,\beta,\gamma\rbrack}(z)
\rightarrow
({{df}\over{dz}})^{n+h^{\lbrack{\alpha,\beta,\gamma}\rbrack}}
B_{n}^{\lbrack\alpha_n,\beta_n,\gamma_n\rbrack}e^{\lbrack\alpha,\beta,\gamma\rbrack}
(f(z))B_{n-k+1}^{\lbrack\alpha_n,\beta_n,\gamma_n\rbrack}e^{\lbrack\alpha,\beta,\gamma\rbrack}(f(z))
\cr
+
\sum_{k=2}^{n+1}
{{n!}\over{k!(n-k+1)!}}
({{df}\over{dz}})^{n-k+1+h^{\lbrack{\alpha,\beta,\gamma}\rbrack}}
B_{k-1}(\lambda(k,n,h^{\lbrack{\alpha_n,\beta_n,\gamma_n}\rbrack})log(f^\prime(z)))
\cr\times
B_{n-k+1}^{\lbrack\alpha_n,\beta_n,\gamma_n\rbrack}e^{\lbrack\alpha,\beta,\gamma\rbrack}(f(z))
}}
with the weight  factor $\lambda$ given by
\eqn\grav{\eqalign{
\lambda
(k,n,h^{\lbrack{\alpha_n,\beta_n,\gamma_n}\rbrack})
=
{kh^{\lbrack{\alpha_n,\beta_n,\gamma_n}\rbrack}}+n-k+1
\cr
+
\alpha_n^2-\alpha_n\alpha-\beta_n^2+\beta_n\beta-\gamma_n^2+\gamma_n\gamma}}

This defines the global conformal transformations for all the string field components
(9) under $z\rightarrow{f(z)}$.
Finally, to prepare for the computation of the SFT  correlators,
we need to  determine the BRST transformation of the string field (9),
in order to compute $<<\Psi,Q\Psi>>=<\Psi I\circ(Q\Psi)>$
where $I$ is the conformal transformation $z\rightarrow{w=-{1\over{z}}}$.
Since all the components of $\Psi$ carry $b-c$ ghost number $+1$,
it shall be sufficient to compute the terms of $Q\Psi$ carrying the $b-c$ ghost number
$+2$, that is, the commutator of $\Psi$ with the stress tensor part of
$Q$ given by $\oint{{dz}\over{2i\pi}}(cT-b{c}\partial{c})$.
Moreover, since $\Psi$ is pure ghost, it is suffucient to consider
the ghost part of $T(z)$.
Another  simplification stems from the fact that the bosonized expression
for $cT_{b-c}-bc\partial{c}$ (where 
$T_{b-c}={1\over2}(\partial\sigma)^2+{3\over2}\partial^2\sigma$ is the $b-c$ part of the stress-tensor)
given by 
\eqn\lowen{:cT_{b-c}-bc\partial{c}:=
e^\sigma(T_{b-c}(z)-\partial^2\sigma)}
so the effect of the second term is just reducing the background charge by 
1 unit; in particular, in our computations of
$I\circ(Q\Psi)=Q(I\circ\Psi)$ this results in effective shifts of the conformal dimensions
of $e^{\lbrack\alpha_n,\beta_n,\gamma_n\rbrack}$ according to
\eqn\lowen{h^{\lbrack\alpha_n,\beta_n,\gamma_n\rbrack}\rightarrow
{\tilde{h}}^{\lbrack\alpha_n,\beta_n,\gamma_n\rbrack}=h^{\lbrack\alpha_n,\beta_n,\gamma_n\rbrack}
+\gamma_n}
Straightforward computation of $Q(I\circ\Psi)$ then gives:

\eqn\grav{\eqalign{Q(I{\circ}B_{n}^{\lbrack\alpha,\beta,\gamma,\rbrack}e^{\lbrack\alpha,\beta,\gamma\rbrack})(w)|_{w=-{1\over{z}}}
\cr
=\sum_{k=1}^{n+1}{{n!}\over{k!}}
\times\lbrack
(k-\delta_1^k)h^{\lbrack\alpha_n,\beta_n,\gamma_n\rbrack}
+\delta_1^k
h^{\lbrack\alpha,\beta,\gamma,\rbrack}
\cr
+(1-\delta_1^k)(\alpha_n^2-\alpha_n\alpha-\beta_n^2+\beta_n\beta-\gamma_n^2+\gamma_n\gamma)
\rbrack
\cr\times
w^{2(h^{\lbrack\alpha,\beta,\gamma,\rbrack}+n-k+1)}
B_{k-1}(x_1,...x_k)|_{x_j=(-1)^j2(k-1)!w^j;j=1,...,k}
\cr\times
\lbrace
\sum_{l=1}^{n-k+1}\sum_{m=0}^{n-k+1-l}{{(-1)^m}\over{(n-k-l+2)!(l+m)!}}
\cr
\times\lbrack
(l-\delta_1^l){\tilde{h}}^{\lbrack\alpha_n,\beta_n,\gamma_n\rbrack}
+\delta_1^l
{\tilde{h}}^{\lbrack\alpha,\beta,\gamma,\rbrack}
+(1-\delta_1^l)(\alpha_n^2-\alpha_n\alpha-\beta_n^2+\beta_n\beta-\gamma_n^2+\gamma_n\gamma)
\rbrack
\cr
\times
\lbrack
\partial^{l+m}c{B^{n-k-l-m+2|n-k-l+2}_{001|\alpha_n\beta_n\gamma_n}}e^{\lbrack\alpha,\beta,\gamma\rbrack}(w)
\rbrace
\cr
+
\sum_{k=1}^n{{(-1)^k}\over{(k-1)!}}\lbrack
{{{\partial^{k+1}}c}\over{k+1}}{B^{n-k|n}_{001|\alpha_n\beta_n\gamma_n}}
e^{\lbrack\alpha,\beta,\gamma\rbrack}(w)
\cr
+
{{{\partial^{k}}c}\over{k}}\partial({B^{n-k|n}_{001|\alpha_n\beta_n\gamma_n}}
e^{\lbrack\alpha,\beta,\gamma\rbrack}(w))
+c\partial({B_n^{{\lbrack\alpha_n,\beta_n,\gamma_n\rbrack}}e^{\lbrack\alpha,\beta,\gamma\rbrack}})(w)
}}

Here $B^{m|n}_{pqr|\alpha\beta\gamma}$ are the conformal dimension $m$ polynomials in bosonized
ghost fields appearing in the OPE of Bell polynomials with exponential fields,
defined according to:
\eqn\grav{\eqalign{
B_n^{\lbrack\alpha_n,\beta_n,\gamma_n,\rbrack}(z_1)e^{\lbrack{p,q,r}\rbrack}(z_2)
=\sum_{m=0}^n{{{:B^{m|n}_{pqr|\alpha\beta\gamma}}(z_1)e^{\lbrack{p,q,r}\rbrack}(z_2):}\over{(z_1-z_2)^{n-m}}}}}
(note the upper script for $B^{m|n}$  chosen here in order not to
confuse them with the incomplete Bell  polynomials for which the lower indices are
reserved according to (15))
It is straightforward to compute
the manifest expressions for  $B^{m|n}_{pqr|\alpha\beta\gamma}$ in terms of incomplete Bell polynomials.
Using the representation (15) for  the Bell polynomials in terms of partitions,
 computing the OPE (42) and extracting the relevant coefficients we get
\eqn\grav{\eqalign{B^{m|n}_{pqr|\alpha\beta\gamma}
={{(-1)^{n-m}n!}\over{(n-m)!m!}}\sum_{k=1}^n\sum_{l=max(1;k-m)}^{min(n-m;k)}
B_{n-m|l}(0!,1!,...,(n-m-l)!)B_{m|k-l}^{\lbrack\alpha,\beta,\gamma\rbrack} 
}}
Here $B_{m|k-l}^{\lbrack\alpha,\beta,\gamma\rbrack}$ are the incomplete Bell polynomials 
in the ghost fields defined according to (10), (14).
The numerical coefficients
$B_{n-m|l}(0!,1!,...,(n-m-l)!)$ given by the values of incomplete Bell polynomials
$B_{n-m|l}(x_1,...x_{n-m-l+1})$ at $x_j=(j-1)!(j=1,...;n-m-l+1)$
and coincide  with
$(n-m)$'th order expansion coefficients of $log^l(1+x)$ around $x=0$.

\centerline{\bf 3. Computation of the Star Product}

Finally, before starting the computation of the correlators, we shall comment
on cohomology constraints on $\Psi^{(+)}$ and $\Psi^{(-)}$, playing the
role of fixing the gauge in SFT and  defining the limits of summation
over $n$ in (9).Since the analytic solution we are looking for, is aiming to describe the higher
spin vacuum, and the higher spin algebras are determined by the structure  of OPE's between
ghost cohomologies $H_{n}\sim{H_{-n-2}}(n>0)$, we impose the following weak cohomology constraints
according to:
\eqn\grav{\eqalign{\Gamma(z)\Psi^{(-)}(w)\sim{O(z-w)^0}
\cr
\Gamma(z)\Psi^{(-)}\approx{0}
}}
The first of these conditions ensures that $\Psi^{(-)}$ is nonsingular under the picture-changing
transformation; the second constraint requires that $\Psi^{(-)}$ is annihilated by $\Gamma$ in
a weak sense, that is, up to terms not contributing to the correlation functions we are considering.
Technically, this implies that, with the picture-changing operator:
\eqn\grav{\eqalign{
\Gamma=-{1\over2}e^\phi\psi_m\partial{X^m}+{1\over4}be^{2\phi-\chi}(\partial\chi+\partial\sigma)
+ce^\chi\partial\chi}}
$\Psi^{(-)}$ is annihilated by the first two terms of $\Gamma$ but is allowed to have
a nonvanishing nonsingular OPE with the last one. However, as the transformation by $c\partial\xi$ shifts the
$b-c$ and $\xi-\eta$ ghost numbers of the string field components by 1 unit,
the terms obtained as a result of the picture-changing will not contribute to the correlators.
Since positive and negative cohomologies are isomorphic, it is sufficient
to consider the constraints on $\Psi^{(-)}$
The constraints (44) ensure that string field components components
of different ghost numbers are unrelated by picture changing (up to terms irrelevant for
correlators). This leads to 
\eqn\grav{\eqalign{
\beta_n=0\cr
\gamma_n=0;1\cr
n\leq{N-1}
}} 
With the constraints (44) $\Psi^{(+)}$ is also automatically annihilated  by the 
inverse picture changing operator $\Gamma^{-1}$, at least in the weak sense.
This condition is stronger than standard gauge constraints on string fields.
 Note that the vanishing of $\beta_n$ particularly ensures
the standard gauge condition $\eta_0\Psi=0$. This is the condition typically imposed on
string fields  at particular fixed ghost number. In our case, however, the constraints
clearly have to be stronger than that since we allowed the contributions from all the ghost numbers.
Note  that the condition $\beta_n=0$ also technically reduces 
the string field (9)
to the  small Hilbert space (although
for generic $\alpha_n,\beta_n,\gamma_n$ $\Psi$ belongs to the large space).
Indeed, the only $\chi$-dependence of $\Psi$ is the 
common $e^\chi$-factor for the comnponents of $\Psi^{(+)}$.
But this factor simply ensures the cancellation of the $\chi$-ghost's background
charge and does not affect the rest of the calculations 
which are effectively in the small space.
Also, in our calculation of the correlators we fix the gauge $\alpha_n=n$. 
This is done for simplicity 
of our calculations; it is  straightforward to generalize them for arbitrary $\alpha_n$.
Finally, we find that the $\gamma_n=1$ choice in (46) is the only one leading
to nontrivial recursion relation on $\lambda_N^n$-coefficients in (11)
with the opposite
 the $\gamma_n=0$ choice trivializing the correlators and destroying the general structure 
of the solutions.
With $I\circ{Q\Psi}$ determined, we are now prepared to calculate
$<<\Psi,Q\Psi>>$ for the string field (9).

To compute the correlators, the following operator products are of importance:

\eqn\grav{\eqalign{
B_n^{\lbrack{\alpha,\beta,\gamma}\rbrack}(z)e^{\lbrack{p,q,r}\rbrack}(w)
=\sum_{k=1}^n\sum_{l=o}^{k}\sum_{m=l}^{n-k+l}(z-w)^{-m}
\cr
\times
{{n!}\over{m!(n-m)!}}B_{m|l}(0!,1!,...(m-l)!)
:B_{n-m|k-l}^{\lbrack{\alpha,\beta,\gamma}\rbrack}(z)e^{\lbrack{p,q,r}\rbrack}(w):
}}

\eqn\grav{\eqalign{
B_{n_1|k_1}^{\lbrack{\alpha,\beta,\gamma}\rbrack}(z)B_{n_2|k_2}^{\lbrack{p,q,r}\rbrack}(w)
=\sum_{l=0}^{min(k_1,k_2)}\sum_{m_1=l}^{n_1-k_1+l}\sum_{m_2=l}^{n_2-k_2+l}
(z-w)^{-m_1-m_2}{{n_1!n_2!}\over{(n_1-m_1)!(n_2-m_2)!}}
\cr
\Lambda_{Bell}(m_1,m_2|l)
:B_{n_1-m_1|k_1-l}^{\lbrack{\alpha,\beta,\gamma}\rbrack}(z)B_{n_2-m_2|k_2-l}^{\lbrack{p,q,r}\rbrack}:
(w)
}}

where the signs  of the normal ordering imply the absence of contractions on the 
right hand side and
the generalized  Bell numbers 

$\Lambda_{Bell}(m_1,m_2|l)$
 are defined as follows.
Let $0<p_1\leq{p_2}...{\leq}p_l$
and 
$0<q_1\leq{p_2}...{\leq}q_l$

be the ordered length $l$ partitions of $m_1$ and $m_2$.
Then
\eqn\grav{\eqalign{
\Lambda_{Bell}(m_1,m_2|l)
=m_1!m_2!\sum_{m_1|p_1,...,p_l}^{partitions}\sum_{m_2|q_1,...,q_l}^{partitions}
\cr
\sum^{pairings}_{p_{i_k};q_{j_k};i_k,j_k=1,...,l}
{{(p_{i_1}+q_{j_1}-1)!...(p_{i_l}+q_{j_l}-1)!}\over
{p_1!...p_l!q_1!...q_l!r_{p_1}!...r_{p_l}!r_{q_1}!...r_{q_l}!}}
}}

where

$r_{p,q}$ are multiplicities of
$p$ and $q$ entering the partitions.

The ghost number anomaly
cancellation condition requires that each correlator
(both 2-point and 3-point)
must have $b-c$ ghost number 3, $\phi$-ghost number $-2$
and  $\chi$-ghost number $1$ .
It is this condition that ensures the triangular
form of (11), making it a well-defined recurrence relation.

The
straightforward calculation of $<<\Psi,Q\Psi>>$ gives

\eqn\grav{\eqalign{
<<\Psi,Q\Psi>>
\cr
=\sum_{N=1}^\infty\sum_{n=0}^{N-1}(\lambda_N^n)^2
\sum_{k=1}^{n+1}\sum_{l=1}^{n-k+1}\sum_{m=0}^{n-k-l}
\sum_{L_1=0}^n
{{(n!)^2}\over{(n-k-l-m+2)!(l+m-1)!(n-L_1)!}}
\cr
\lbrace
\lbrack
(k-\delta_1^k){h^{\lbrack{n,0,0}\rbrack}}
+\delta_1^k{h^{\lbrack{-(N+2),0,1}\rbrack}}
+(1-\delta_1^k)((N+2)(n+1)-1)
\rbrack
\cr
\times\lbrack
(l-\delta_1^l){{\tilde{h}}^{\lbrack{n,0,0}\rbrack}}
+\delta_1^l{{\tilde{h}}^{\lbrack{-(N+2),0,1}\rbrack}}
+(1-\delta_1^l)((N+2)(n+1)-1)\rbrack
\cr
\times
\sum_{k_1=1}^n\sum_{k_2=1}^{l+m-1}\sum_{k_3=1}^{n-k-l+2}
\sum_{l_1=1}^{min(L_1;k_1-1)}\sum_{{l_2}=1}^{min(m,k_3-1)}
(-1)^{k+l_1+l_2+L_1}
\cr
(2+n(N+2))^{l_1}
B_{L_1|l_1}(0!,...,(L_1-l_1)!)B_{L_2|l_2}(0!,...,(L_2-l_2)!)
\cr
\times
\sum_{q=1}^{k_3-l_3}\sum_{M=1}^{n+2-k-l-m-k_3+l_2-q}
(nN-1)^qB_{M|q}(0!,...,(M-q)!)
\cr
\times
\sum_{Q=k_3-l_2-q}^{n-L_1-k_1-l_1-k_3+l_2+q}{{(-1)^{Q+n-L_1}(n-L_1)!}\over{Q!(n-L_1-Q)!}}
\Lambda_{Bell}(Q;n+2-k-l-m-M|k_3-l_3-q)
\cr
\times
\lbrack
n^{k_2}\delta_{k_2}^{k_1-k_3-l_1+l_2+q}
\Lambda_{Bell}(n-L_1-Q;l+m-1|k_2)
\cr
-(l+m-1)n^{k_2-1}\delta_{k_2-1}^{k_1-k_3-l_1+l_2+q}
\Lambda_{Bell}(n-L_1-Q;l+m-2|k_2)\rbrack
\rbrace
}}

The next step is to compute the $3$-point correlator
$$<<\Psi,\Psi\star\Psi=<h\circ{f_1^3}\circ\Psi(0)h\circ{f_2^3}\circ\Psi(0)
h\circ{f_3^3}\circ\Psi(0)>$$
where, for the convenience of the computation 
the conformal transformation 
\eqn\grav{\eqalign{h(z)=-i{{z-1}\over{z+1}}}}
further maps the disc to the half-plane
(upon mapping the worldsheets of 3 interacting strings  to the disc).
The straightforward  computation  of the $3$-point function,
using the operator products (47) ,(48) gives
\eqn\grav{\eqalign{<<\Psi,\Psi\star\Psi>>
=
\cr
\sum_{N=1}^\infty\sum_{N_1,N_2=1}^{N_1+N_2=N-2}\lambda_N^n\lambda_{N_1}^{n_1}\lambda_{N_2}^{n_2}
\sum_{k=1}^{n+1}\sum_{k_1=1}^{n_1+1}\sum_{k_2=1}^{n_2+1}
{{n!n_1!n_2!}\over{(n-k+1)!(n_1-k_1+1)!(n_2-k_2+1)!k!k_1!k_2!}}
\cr
\times
\lbrack
(k-\delta_1^k)h^{\lbrack{n,0,1}\rbrack}
+\delta_1^kh^{\lbrack{N,0,1}\rbrack}
+(1-\delta_1^{k})(n^2-nN)
\rbrack
\cr
\times
\lbrack
(k_1-\delta_1^{k_1})h^{\lbrack{n_1,0,1}\rbrack}
+\delta_1^kh^{\lbrack{-N_1-2,0,1}\rbrack}
+(1-\delta_1^{k_1})(n_1^2+n_1(N_1+2))
\rbrack
\cr
\times
\lbrack
(k_2-\delta_1^{k_2})h^{\lbrack{n_2,0,1}\rbrack}
+\delta_1^{k_2}h^{\lbrack{-N_2-2,0,1}\rbrack}
+(1-\delta_1^{k_2})(n_2^2+n_2(N_2+2))
\rbrack
\cr
({2\over3})^{h^{\lbrack{N,1,1}\rbrack}+h^{\lbrack{-N_1-2,0,1}\rbrack}+h^{\lbrack{-N_2-2,0,1}\rbrack}+n+n_1+n_2-k-k_1-k_2}
\cr\times
B_{k-1}(\lambda(k,n,h^{\lbrack{n,0,1}\rbrack})log(h{\circ}(f_1^3)^\prime(z)))|_{z=0}
\cr
B_{k_1-1}(\lambda(k_1,n_1,h^{\lbrack{n_1,0,1}\rbrack})log(h{\circ}(f_2^3)^\prime(z)))|_{z=0}
B_{k_2-1}(\lambda(k_2,n_2,h^{\lbrack{n_2,0,1}\rbrack})log(h{\circ}(f_3^3)^\prime(z)))|_{z=0}
\cr
\times
\sum_{m=1}^{n-k+1}\sum_{m_1=1}^{(n_1-k_1+1)}\sum_{m_2=1}^{(n_2-k_2+1)}
\sum_{s_1=0}^{m}\sum_{s_2=0}^{m-s_1}\sum_{t_1=0}^{m_1}\sum_{t_2=0}^{m_1-t_1}
\sum_{u_1=0}^{m_2}\sum_{u_2=0}^{m_2-u_1}
\cr
\sum_{L_1=s_1}^{(n-k+1-m+s_1)}\sum_{L_2=s_2}^{(n-k+1-m+s_2-L_1)}
\sum_{M_1=t_1}^{(n_1-k_1+1-m_1+t_1)}
\cr
\sum_{M_2=t_2}^{(n_1-k_1+1-m_1+t_2-M_1)}
\sum_{P_1=u_1}^{(n_2-k_2+1-m_2+u_1)}
\sum_{P_2=u_2}^{(n_2-k_2+1-m_2+u_2-P_1)}
\cr
\lbrace
B_{L_1|s_1}(0!,1!,...,(L_1-s_1)!)B_{L_2|s_2}(0!,1!,...,(L_2-s_2)!)
B_{M_1|t_1}(0!,1!,...,(M_1-t_1)!)
\cr
B_{M_2|t_2}(0!,1!,...,(M_2-t_2)!)
B_{P_1|u_1}(0!,1!,...,(P_1-u_1)!)B_{P_2|u_2}(0!,1!,...,(P_2-u_2)!)
\cr
(-(N_1+2)n-1)^{L_1}(-(N_2+2)n-1)^{L_2}(-n_1N+1)^{M_1}
\cr
(-n_1(N_2+2)-1)^{M_2}
(-n_2N-1)^{P_1}(n_2(N_1+2)+1)^{P_2}
\cr
({\sqrt{3}})^{N(N_1+2)-(N_1+2)(N_2+1)+2-L_1-M_1-M_2-P_2}
(2{\sqrt{3}})^{N(N_2+2)+1-L_2-P_1}
\rbrace
\cr
\times
(-nn_1+1)^{r_1}
(-nn_2+1)^{r_2}
(-n_1n_2+1)^{r_3}
\sum_{R_1=r_1}^{n-k+1+r_1}
\sum_{R_2=r_2}^{n_1-k_1+1+r_2}
\sum_{R_3=r_3}^{n_2-k_2+1+r_3}
\cr
({\sqrt{3}})^{-R_1+R_3}(2{\sqrt{3}})^{-R_1-R_3}
\Lambda_{Bell}(R_1;R_2|r_1)
\Lambda_{Bell}(R_1;R_3|r_2)
\Lambda_{Bell}(R_2;R_3|r_3)
\rbrace}}

where

\eqn\grav{\eqalign{
r_1={1\over2}(m+m_1-m_2-s_1-s_2-t_1-t_2+u_1+u_2)\cr
r_2={1\over2}(m-m_1+m_2-s_1-s_2+t_1+t_2-u_1-u_2)\cr
r_1={1\over2}(-m+m_1-m_2+s_1+s_2-t_1-t_2+u_1+u_2)}}

The values of the complete Bell polynomials
appearing as  a result of conformal transformations  by  $h\circ{f_j^3}(j=1,2,3)$
in the cubic term (52), are calculated  
 to be given by
\eqn\grav{\eqalign{B_{k}(\lambda(k,n,h^{\lbrack{n,0,1}\rbrack})log(h{\circ}(f_j^3)^\prime(z)))|_{z=0}
=
\sum_{p=0}^k\sum_{l=0}^{p}\sum_{m=0}^{k-p}
{{{i^{p+1}}(-1)^{l+m}k!}\over{l!(p-l)!m!(k-p-m)!
}}
\cr\times
{{\Gamma(\lambda-{2\over3})\Gamma(\lambda+{2\over3})(\Gamma(\lambda+1))^2}\over{
\Gamma(\lambda-{2\over3}-l)
\Gamma(\lambda+{2\over3}+l-k)\Gamma(\lambda+1-m)\Gamma(\lambda+1+m+p-k)}}
}}
Combining our results for two-point and three-point correlators (50)-(52) we deduce the following
recurrence relation for the $\lambda_N^n$ structure constants entering the analytic solutions:
\eqn\grav{\eqalign{
\rho_{N|LN_1;N_2}^{n|n_1;n_2}={{{(\kappa_3)}_{N|N_1;N_2}^{n|n_1;n_2}}\over
{{(\kappa_2)}_{N}^{n}
}}}}
with

\eqn\grav{\eqalign{
{(\kappa_2)}_{N}^{n}=
\sum_{k=1}^{n+1}\sum_{l=1}^{n-k+1}\sum_{m=0}^{n-k-l}
\sum_{L_1=0}^n
{{(n!)^2}\over{(n-k-l-m+2)!(l+m-1)!(n-L_1)!}}
\cr
\lbrace
\lbrack
(k-\delta_1^k){h^{\lbrack{n,0,0}\rbrack}}
+\delta_1^k{h^{\lbrack{-(N+2),0,1}\rbrack}}
+(1-\delta_1^k)((N+2)(n+1)-1)
\rbrack
\cr
\times\lbrack
(l-\delta_1^l){{\tilde{h}}^{\lbrack{n,0,0}\rbrack}}
+\delta_1^l{{\tilde{h}}^{\lbrack{-(N+2),0,1}\rbrack}}
+(1-\delta_1^l)((N+2)(n+1)-1)\rbrack
\cr
\times
\sum_{k_1=1}^n\sum_{k_2=1}^{l+m-1}\sum_{k_3=1}^{n-k-l+2}
\sum_{l_1=1}^{min(L_1;k_1-1)}\sum_{{l_2}=1}^{min(m,k_3-1)}
(-1)^{k+l_1+l_2+L_1}
\cr
(2+n(N+2))^{l_1}
B_{L_1|l_1}(0!,...,(L_1-l_1)!)B_{L_2|l_2}(0!,...,(L_2-l_2)!)
\cr
\times
\sum_{q=1}^{k_3-l_3}\sum_{M=1}^{n+2-k-l-m-k_3+l_2-q}
(nN-1)^qB_{M|q}(0!,...,(M-q)!)
\cr
\times
\sum_{Q=k_3-l_2-q}^{n-L_1-k_1-l_1-k_3+l_2+q}{{(-1)^{Q+n-L_1}(n-L_1)!}\over{Q!(n-L_1-Q)!}}
\Lambda_{Bell}(Q;n+2-k-l-m-M|k_3-l_3-q)
\cr
\times
\lbrack
n^{k_2}\delta_{k_2}^{k_1-k_3-l_1+l_2+q}
\Lambda_{Bell}(n-L_1-Q;l+m-1|k_2)
\cr
-(l+m-1)n^{k_2-1}\delta_{k_2-1}^{k_1-k_3-l_1+l_2+q}
\Lambda_{Bell}(n-L_1-Q;l+m-2|k_2)\rbrack
\rbrace
}}
and
\eqn\grav{\eqalign{
{(\kappa_3)}_{N|N_1;N_2}^{n|n_1;n_2}=
\sum_{k=1}^{n+1}\sum_{k_1=1}^{n_1+1}\sum_{k_2=1}^{n_2+1}
{{n!n_1!n_2!}\over{(n-k+1)!(n_1-k_1+1)!(n_2-k_2+1)!k!k_1!k_2!}}
\cr
\times
\lbrack
(k-\delta_1^k)h^{\lbrack{n,0,1}\rbrack}
+\delta_1^kh^{\lbrack{N,0,1}\rbrack}
+(1-\delta_1^{k})(n^2-nN)
\rbrack
\cr
\times
\lbrack
(k_1-\delta_1^{k_1})h^{\lbrack{n_1,0,1}\rbrack}
+\delta_1^kh^{\lbrack{-N_1-2,0,1}\rbrack}
+(1-\delta_1^{k_1})(n_1^2+n_1(N_1+2))
\rbrack
\cr
\times
\lbrack
(k_2-\delta_1^{k_2})h^{\lbrack{n_2,0,1}\rbrack}
+\delta_1^{k_2}h^{\lbrack{-N_2-2,0,1}\rbrack}
+(1-\delta_1^{k_2})(n_2^2+n_2(N_2+2))
\rbrack
\cr
({2\over3})^{h^{\lbrack{N,1,1}\rbrack}+h^{\lbrack{-N_1-2,0,1}\rbrack}+h^{\lbrack{-N_2-2,0,1}\rbrack}+n+n_1+n_2-k-k_1-k_2}
\cr\times
B_{k-1}(\lambda(k,n,h^{\lbrack{n,0,1}\rbrack})log(h{\circ}(f_1^3)^\prime(z)))|_{z=0}
\cr
B_{k_1-1}(\lambda(k_1,n_1,h^{\lbrack{n_1,0,1}\rbrack})log(h{\circ}(f_2^3)^\prime(z)))|_{z=0}
B_{k_2-1}(\lambda(k_2,n_2,h^{\lbrack{n_2,0,1}\rbrack})log(h{\circ}(f_3^3)^\prime(z)))|_{z=0}
\cr
\times
\sum_{m=1}^{n-k+1}\sum_{m_1=1}^{(n_1-k_1+1)}\sum_{m_2=1}^{(n_2-k_2+1)}
\sum_{s_1=0}^{m}\sum_{s_2=0}^{m-s_1}\sum_{t_1=0}^{m_1}\sum_{t_2=0}^{m_1-t_1}
\sum_{u_1=0}^{m_2}\sum_{u_2=0}^{m_2-u_1}
\cr
\sum_{L_1=s_1}^{(n-k+1-m+s_1)}\sum_{L_2=s_2}^{(n-k+1-m+s_2-L_1)}
\sum_{M_1=t_1}^{(n_1-k_1+1-m_1+t_1)}
\cr
\sum_{M_2=t_2}^{(n_1-k_1+1-m_1+t_2-M_1)}
\sum_{P_1=u_1}^{(n_2-k_2+1-m_2+u_1)}
\sum_{P_2=u_2}^{(n_2-k_2+1-m_2+u_2-P_1)}
\cr
\lbrace
B_{L_1|s_1}(0!,1!,...,(L_1-s_1)!)B_{L_2|s_2}(0!,1!,...,(L_2-s_2)!)
B_{M_1|t_1}(0!,1!,...,(M_1-t_1)!)
\cr
B_{M_2|t_2}(0!,1!,...,(M_2-t_2)!)
B_{P_1|u_1}(0!,1!,...,(P_1-u_1)!)B_{P_2|u_2}(0!,1!,...,(P_2-u_2)!)
\cr
(-(N_1+2)n-1)^{L_1}(-(N_2+2)n-1)^{L_2}(-n_1N+1)^{M_1}
\cr
(-n_1(N_2+2)-1)^{M_2}
(-n_2N-1)^{P_1}(n_2(N_1+2)+1)^{P_2}
\cr
({\sqrt{3}})^{N(N_1+2)-(N_1+2)(N_2+1)+2-L_1-M_1-M_2-P_2}
(2{\sqrt{3}})^{N(N_2+2)+1-L_2-P_1}
\rbrace
\cr
\times
(-nn_1+1)^{r_1}
(-nn_2+1)^{r_2}
(-n_1n_2+1)^{r_3}
\sum_{R_1=r_1}^{n-k+1+r_1}
\sum_{R_2=r_2}^{n_1-k_1+1+r_2}
\sum_{R_3=r_3}^{n_2-k_2+1+r_3}
\cr
({\sqrt{3}})^{-R_1+R_3}(2{\sqrt{3}})^{-R_1-R_3}
\Lambda_{Bell}(R_1;R_2|r_1)
\Lambda_{Bell}(R_1;R_3|r_2)
\Lambda_{Bell}(R_2;R_3|r_3)
\rbrace}}

This concludes the computation of the coefficients defining the analytic  OSFT
solution.

\centerline{\bf 4. Bell Polynomials and Higher Spin Algebras}

In this concluding section we shall present some arguments relating the 
structure of the analytic OSFT solution, studied in this work,
to free field realizations of the higher spin algebras in $AdS$.
An
insightful hint, relating Bell polynomials
to  free field realizations of higher spin algebras
in AdS, comes from $c=1$ model, i.e. one-dimensional
noncritical string theory.

The one-dimensional string 
compactified on $S^1$
has no standard massless modes
(like a photon) but does have a $SU(2)$ multiplet of massless
states existing at nonstandard $b-c$ ghost numbers and discrete mumentum values
~{\discf,\discs, \ias}

The $SU(2)$ symmetry at self-dual radius $R={1\over{{\sqrt{2}}}}$
is realized by the operators:
\eqn\grav{\eqalign{
T_{\pm}=\oint{dz}e^{{\pm}iX{\sqrt{2}}};T_0=\oint{dz}\partial{X}
}}

The SU(2) multiplet of {discrete states}
can be constructed  by acting with the lowering $T_{-}$ of
$SU(2)$ on the highest weight vectors given by
tachyonic  primaries

$V_l=e^{(ilX+(l-1)\varphi){\sqrt{2}}}$:
(with integer l)
\eqn\grav{\eqalign{U_{l|m}=T_{-}^{l-m}V_l
}}
Manifest expressions for {$U_{l|m}$} vertex operators are complicated,
however, their structure constants have been  deduced by ~{\discf, \discs}
 by using symmetry arguments.
One has
\eqn\grav{\eqalign{
U_{l_1|m_1}(z)U_{l_2|m_2}(w)\sim{(z-w)^{-1}}{C(l_1,l_2,l_3|m_1,m_2,m_3)}f(l_1,l_2)U_{l_3,m_3}
}}
where the $SU(2)$ Clebsch-Gordan coefficients are fixed by the symmetry while
the function of Casimir eigenvalues $f(l_1,l_2)$ is nontrivial and was deduced to be given by
\eqn\grav{\eqalign{
f(l_1,l_2)
={{{\sqrt{l_1+l_2}}(2l_1+2l_2-2)!}\over{{\sqrt{2l_1l_2}}(2l_1-1)!(2l_2-1)!}}
}}

~{\discf, \discs}

Remarkably, these structure constants coincide exactly
with those of higher spin algebra in $AdS_3$ in a certain basis,
computed by E. Fradkin and V. Linetsky in 1989, in  a seemingly
different context ~{\disct, \discft}
On the other hand, the explicit structure of the vertex operators
for the discrete states realizing this algebra is given by

\eqn\grav{\eqalign{
U_{l|m}\sim
\sum_{{1\over2}(l(l-1)-m(m-1))|p_1,...,p_{l-m}}
B_{p_1}(-iX{\sqrt{2}})...B_{p_{l-m}}(-iX{\sqrt{2}})e^{{\sqrt{2}}(imX+(l-1)\varphi)}
}}

with the sum taken over ordered partitions of
${1\over2}(l(l-1)-m(m-1))|p_1,...,p_{l-m}$

This is a relatively simple example
 of Bell polynomials multiplied by exponentials
realizing the higher spin algebras in $AdS_{d}$
in terms of vertex operator algebras in $d-1$-dimensional
string theory.
One can further think of the extending the symmetry
of the $c=1$ model by supersymmetrizing it on the worldsheet
and coupling to the $\beta-\gamma$ enhancing the
symmetry from $SU(2)$ to $SU(4)$
The SU(4) algebra can be realized by taking the
raising generators:

\eqn\grav{\eqalign{
T{0;1}=\oint{dz}{e^{iX}}\psi
T^{-3;2}=\oint{dz}{e^{-3\phi+2iX}}\psi
\cr
T^{-4;3}=
\oint{dz}{e^{-4\phi+3iX}}\psi}}

and the remaining 12 generators are obtained by
acting on (63) with the lowering generators of $SU(2)$,
$T^{0;-1}=\oint{dz}e^{-iX}\psi$:

\eqn\grav{\eqalign{
T^{-4;k}=(T^{0;-1})^{3-k}T^{-4;3}\cr
T^{-3;l}=(T^{0;-1})^{2-k}T^{-3;2}\cr
T^{0;m}=(T^{0;-1})^{1-k}T^{0;1}\cr
-3\leq{k}\leq{3}\cr
-2\leq{l}\leq{2}\cr
-1\leq{m}\leq{1}}}

with $T^{0;0},T^{-3;0}.T^{-4;0}$ being the Cartan generators
of $SU(4)$.
The $SU(4)$ multiplet is then obtained by 
the combinations of lowering operators  of $SU(4)$
with negative $k,l,m$
 acting on the same dressed tachyonic primaries entering (59).
The structure constants of the operators of the $SU(4)$  are
again given by the Klebsch-Gordan coefficients of $SU(4)$ multiplied by
certain functions of the eigenvalues of  SU(4) Casimir operators.
Unfortunately, because of the complexity of the SU(4) operators,
the explicit form of these functions has never been worked out.
Evaluating them would be emportant in order to point
out the relation of this operator algebra to the structure constants
of the higher spin algebra in $AdS_5$. With the techniques , explained
in this paper, we hope to be able to perform this computation in the
future work. As in the $SU(2)$ case, the structure of the manifest vertex
operator expressions for the $SU(4)$ multiplet involve  products
of Bell polynomials in $X,\psi$ and ghost fields multiplies
by exponents of $X$, although the structure of the Bell polynomials
in the $SU(4)$ case is  clearly more complicated than (62).
 Thus one can think of Bell polynomial products
(multiplied by exponents) as of  natural vertex operator realizations
of various $AdS$ higher spin algebras in string theory.
The string field theory analytic solution presented in this paper 
is  a simple example of this class of the operators.
By itself, it is clearly incomplete to describe the full higher spin vacuum,
despite some of  its attractive properties.
In particular, one clearly
has to generalize this string field theory solution to include
products of multiple Bell polynomials, in order to make connections to the
full higher spin algebras in $AdS$.
Our particular conjecture is that the 
OSFT
solutions of the type:
\eqn\grav{\eqalign{
\Psi=\sum_{N,n_1,...,n_k}\lambda_N^{n_1...n_k}B_{n_1}(\phi,\chi,\sigma)
...B_{n_k}(\phi,\chi,\sigma)({c\xi{e^{N\phi}}+ce^{-(N+2)\phi}})
}}
can be related to contributions of the
$k$-row higher spin fields with mixed symmetries to the collective
higher  spin vacuum configuration.
In general, the full space of these solutions would form an ``enveloping''
of higher-spin algebra. 
It would be interesting to point out the connection of this enveloping
to multiparticle extensions of the higher spin algebras proposed by
Vasiliev ~{\vmulti}. With the on-shell arguments, showing the relevance
of the vertex operators for the frame-like higher spin fields to 
the background independence, relating the languages and concepts of
higher spin gauge theories and string field theory promises a fascinating
ground for the future work.

\centerline{\bf Acknowledgements}

It is a pleasure to  thank Loriano Bonora
and other organizers and participants 
of SFT-2014 conference at SISSA, Trieste for
hospitality and productive discussions.

This work was partially supported by the National 
Research Foundation of Korea(NRF) grant funded 
by the Korea government(MEST) through the Center for
 Quantum Spacetime(CQUeST) of 
Sogang University with grant number 2005-0049409.
I also acknowledge the support of the NRF grant number 2012-004581.

\listrefs

\end